\documentclass[twocolumn,aps,showpacs,superscriptaddress,amsmath,amsfonts,amssymb,floatfix,pre]{revtex4}
\usepackage{graphicx}
\usepackage{dcolumn}
\usepackage{color}
\usepackage{bm}

\begin{document}

\title{Random matrix analysis of localization properties of Gene co-expression network}
\author{Sarika Jalan}
\affiliation{Department of Physics and Centre for Computational Science and Engineering, National University of Singapore, 117456, Republic of Singapore}
\author{Norbert Solymosi}
\affiliation{Department of the Physics of Complex Systems, E\"otv\"os University,
H-1117 P\'azm\'any P\'eter s\'et\'any 1/A, Budapest, Hungary}
\author{G{\'a}bor Vattay}
\affiliation{Department of the Physics of Complex Systems, E\"otv\"os University, H-1117 P\'azm\'any P\'eter s\'et\'any 1/A, Budapest, Hungary}
\author{ Baowen Li}
\affiliation{Department of Physics and Centre for Computational Science and Engineering, National University of Singapore, 117456, Republic of Singapore}
\affiliation{NUS Graduate School for Integrative Sciences and Engineering, 117546, Republic of Singapore}

\begin{abstract}
We analyze gene co-expression network under the random matrix theory framework. The nearest neighbor spacing distribution 
of the adjacency matrix of this network follows Gaussian orthogonal statistics of random matrix theory (RMT). Spectral 
rigidity test follows random matrix prediction for a certain range, and deviates after wards. Eigenvector analysis of the 
network using inverse participation ratio (IPR) suggests that the statistics of bulk of the eigenvalues of network is 
consistent with those of the real symmetric random matrix, whereas few eigenvalues are localized. Based on these IPR 
calculations, we can divide eigenvalues in three sets; (A) The non-degenerate part that follows RMT. (B) The 
non-degenerate part, at both ends and at intermediate eigenvalues, which deviate from RMT and expected to contain 
information about {\it important nodes} in the network. (C) The degenerate part with $zero$ eigenvalue, which fluctuates 
around RMT predicted value. We identify nodes corresponding to the dominant modes of the corresponding eigenvectors and 
analyze their structural properties.
\end{abstract}


\maketitle

\section{Introduction}
\subsection{Complex Networks}
Gene expression information captured in microarrays data for a variety of environmental and genetic perturbations, in 
conjunction with other sources such as protein-protein/protein-DNA interaction and operon organization data, promises to 
yield unprecedented insights into the organization and functioning of biological systems \cite{p-p1,lit}. It has been 
increasingly realized that dissecting the genetic and chemical circuitry prevents us from further understanding the 
biological processes as a whole. In order to understand the complexities involved, all reactions and processes should be 
analyzed together. To this end, network theory will be used. It has been getting fast recognition to study systems which 
could be defined in terms of units and interactions among them. These studies revealed that the available data from gene 
co-expression network share some unexpected features with other complex networks as diverse as the Internet routers. In 
order to understand the behavior of complex systems such as gene co-expression network, several simple models, based on 
the simple principles and captures some essential features of the system, have been introduced, these models 
are\cite{BA,rev-Strogatz,rev-network}.

In this paper, by using network theory and random matrix theory (RMT), we analyze gene co-expression network. First we 
generate network from the gene co-expression data collected form six brain regions that are metabolically relevant to 
Alzheimer's disease \cite{data_gene} by using appropriate threshold, and then study the spectra of this network under the 
RMT framework. Information about the genes that are preferentially expressed during the course of Alzheimer's disease 
could improve our understanding of the molecular mechanisms involved in the pathogenesis of this common cause of cognitive 
impairment in senior persons, provide new opportunities in the diagnosis, early detection, and tracking of this disorder, 
and provide novel targets for the discovery of interventions to treat and prevent this disorder. Information about the 
genes that are preferentially expressed in relationship to normal neurological aging could provide new information about 
the molecular mechanisms that are involved in normal age-related cognitive decline and a host of age-related neurological 
disorders, and they could provide novel targets for the discovery of interventions to mitigate some of these deleterious 
effects.

Co-expression networks have also been known as relevance networks. The terminology has been introduced by Butte and Kohane 
\cite{Butte}. Since then properties of the relevance networks have been extensively studied \cite{rev-Diogo}.

The paper is organized as follows: after introductory sub-section on the relevance of network theory and gene 
co-expression network, we discuss the recent outcome of RMT analysis of complex networks in the following subsection B. 
Main goals of our eigenvector analysis are written in the subsection C. Section II describes the important achievements of 
RMT and explains its various properties we use in our analysis. Section III sheds light on the data and network 
construction. Section IV presents various numerical results. Section V concludes the paper with a discussion on the 
relevance of current analysis, as well suggests future directions.

\subsection{RMT of Network Spectra}
Our previous work \cite{pre2007a} showed that various vastly studied model networks follow random matrix predictions of 
Gaussian orthogonal statistics (GOE) at the level repulsion domain. We demonstrated that nearest neighbor spacing 
distribution (NNSD) of protein-protein interaction network of budding yeast follows RMT prediction as well 
\cite{pre2007a}. This is a promising result which suggests that these networks can be modeled as a random matrix chosen 
from an appropriate ensemble. The universal GOE statistics of eigenvalues fluctuations could be understood as some kind of 
randomness spreading over the protein-protein interaction network and model networks capturing real world properties. 
Recently, covariance matrix of amino acid displacement has been analyzed under RMT framework \cite{rmt_aminoacid_prl2009}. 
The analysis shows that the bulk of eigenvalues follows universal GOE statistics of RMT. In the present paper, we analyze 
gene co-expression network \cite{data_gene} under RMT framework. First we calculate nearest neighbor spacing distribution 
of network spectra, and then perform eigenvector analysis to detect nodes having specific contribution to network.

\subsection{Important nodes and connections} \label{impnode}
It is now well known that various real world systems are scale-free network\cite{BA}. The scale-free nature of networks 
suggests that there exist few nodes with very high degrees. Motivated by this finding they suggested that since these 
nodes are responsible to hold the whole networks and henceforth are the most important ones. Some other analysis (by 
Newman and others) of real-world networks show that complex networks have community or module structure 
\cite{Newman,community}. Modules are the division of network nodes within which the network connections are dense, but 
between which they are sparser. The modularity concept assumes that system functionality can be partitioned into a 
collection of modules and each module performs an identifiable task, separable from the functions of other modules 
\cite{modular}. Analysis of module structure involves {\it betweenness} measure. Betweenness of an edge is defined as the 
number of shortest path between pairs of nodes going through the edge. Betweenness studies of real world networks suggests 
that the nodes connecting the different communities are the most important ones, which has been verified in the metabolic 
networks by Amaral et. al.\cite{modular}.

Above description emphasizes on the importance of nodes depending on their position in the network, as these nodes 
characterize network properties. On the other hand Erd\"os-R\'enyi (ER) and Strogatz-Watts (SW) models emphasize on the 
importance of random connections in the networks. In the ER model any two nodes are connected with probability $p$. One of 
the most interesting property of ER model is the sudden emergence of various global properties, such as, emergence of a 
giant cluster. As $p$ increases, while number of nodes in the graph remains constant, the giant cluster emerges through a 
phase transition \cite{graph}. Further, the SW model shows the small world transition with the fine tuning of number of 
random connections \cite{SW}. Our previous RMT analysis of the spectra of SW model networks \cite{pre2007a} show that at 
the SW transition there is a transition to the {\it spreading of randomness} in the network characterized by the 
correlations between nearest eigenvalues. In the current paper we analyze spectra of the gene co-expression network under 
RMT framework. Particularly we study eigenvectors of the adjacency matrix of this network. The spectra has two parts, one 
part which follows RMT predictions of universal GOE statistics and other part which does not follow RMT prediction. The 
eigenvectors deviating from the RMT prediction provide information about the {\it influential or important nodes} in the 
network.

\section{Random matrix statistics} \label{RMT}
RMT deals with the statistical properties of matrices with independent random entries. To be self-consistent, we give a 
brief introduction of the RMT here, and explain various RMT properties of eigenvector components which we will use in our 
analysis. RMT was initially proposed to explain the statistical properties of nuclear spectra \cite{mehta}. Later this 
theory was successful applied in the study of the spectra of different complex systems such as disordered systems, quantum 
chaotic systems, large complex atoms \cite{rev-rmt}. Recent studies illustrate the usefulness of RMT in understanding the 
statistical properties of the empirical cross-correlation matrices appearing in the study of multivariate time series of 
followings: the price fluctuations in the stock market \cite{rmt-stock}, EEG data of brain \cite{rmt-brain}, variation of 
various atmospheric parameters \cite{rmt-santh}, etc. Recent analysis of complex networks under RMT framework 
\cite{pre2007a,pre2009a,pre2009b,rmt_aminoacid_prl2009} show that various network models and real world network also 
follow universal GOE statistics. Furthermore localization of eigenvectors have also been used to analyze various 
structural and dynamical properties of real and model networks \cite{Menzinger,Baowen}.

In the following, we introduce spacing distribution and $\Delta_3$ statistics of random matrices. We denote the 
eigenvalues of a network by $\lambda_i,\,\,i=1,\dots,N$, where $N$ is size of the network and $\lambda_1 < \lambda_2 < 
\lambda_3 < \dots < \lambda_N$. In order to get universal properties of the fluctuations of eigenvalues, people usually 
unfold the eigenvalues by a transformation $\overline{\lambda}_i = \overline{N} (\lambda_i)$, where $\overline{N}$ is 
averaged integrated eigenvalue density \cite{mehta}. Since we do not have any analytical form for $\overline{N}$, we 
numerically unfold the spectrum by polynomial curve fitting (for elaborate discussion on unfolding, see Ref.\cite{mehta}). 
After unfolding, average spacings is {\it unity}, independent of the system. Using the unfolded spectra, we calculate 
spacings as $s_i=\overline{\lambda}_{i+1}-\overline{\lambda}_i$. NNSD is defined as the probability distribution ($P(s)$) 
of these $s_i$'s. In the case of GOE statistics,\\
\begin{equation}
P(s)=\frac{\pi}{2}s\exp \left(-\frac{\pi s^2}{4}\right)
\label{eq-goe}
\end{equation}

The $\Delta_3$-statistic measures the least-square deviation of the spectral staircase function representing the averaged 
integrated eigenvalue density $\overline{N}(\lambda)$ from the best straight line fitting for a finite interval $L$ of the 
spectrum, i.e.,
\begin{equation}
\Delta_3(L; x) = \frac{1}{L} \min_{a,b} \int_x^{x+L} \,\left[
N(\overline{\lambda}) - a \overline{\lambda} -b \right]^2\,d \overline{\lambda}
\label{eq-delta3}
\end{equation}
where $a$ and $b$ are obtained from a least-square fit. Average over several choices
of $x$ gives the spectral rigidity $\Delta_3(L)$.  For the GOE case,
$\Delta_3(L)$ depends {\it logarithmically} on $L$, i.e.,
\begin{equation}
\Delta_3(L) \sim \frac{1}{\pi^2} \ln L.
\label{eq-delta3-goe}
\end{equation}

The following sub-section explains the properties of eigenvectors of random matrices.

\subsection{Eigenvector analysis} \label{EV}
The distribution of eigenvectors components are studied to obtain system dependent information. Let $u_l^k$ is the $l$th 
component of $k$th eigenvector $u^k$. The eigenvector components of a GOE random matrix are Gaussian distributed random 
variables, for this the distribution of $r=|u_l^k|^2$, in the limit of large matrix dimension, is given by Porter-Thomas 
distribution \cite{casati},i.e.,
\begin{equation}
P(r) = \frac{N}{\sqrt{2\pi r}} \exp\left(\frac{-N r}{2}\right)
\label{PT}
\end{equation}
Shannon entropy for the state whose components are described by the above distribution,
would be given by in large $N$ limit as \cite{casati},
\begin{equation}
H_s \sim -N \int_0^{\infty} r \ln(r) P(r) dr \sim \ln\left(\frac{N}{2}\right).
\label{eq-entropy}
\end{equation}
Additionally, inverse participation ratio (IPR) is also considered
to study the RMT features of the eigenvectors. The IPR of
eigenvector is defined as
\begin{equation}
I^k = \sum_{l=1}^{N} [u_l^k]^4
\label{eq-IPR}
\end{equation}
where $u_l^k, l=1, \hdots , N$ are the components of eigenvector $u^k$. The meaning of $I^k$ is illustrated by two 
limiting cases : (i) a vector with identical components $u_l^k \equiv 1/\sqrt{N}$ has $I^k =1/N$, whereas (ii) a vector 
with one component $u_1^k=1$ and the remainders zero has $I^k=1$. Thus, the IPR quantifies the reciprocal of the number of 
eigenvector components that contribute significantly. For a vector with components following distribution (\ref{PT}) has 
$I^k \sim 3/N$.

\section{Data and network construction}
The data-set (GSE5281) was obtained from Gene Expression Omnibus \cite{data_gene}. Liang et al. \cite{lit} studied gene 
expression profiles from laser capture micro dissected neurons in six functionally and anatomically distinct regions from 
clinically and histopathologically normal aged human brains. From these data-sets only $74$ normal samples were used to 
construct the co-expression networks. In the original study the Affymetrix Human Genome U133 Plus 2.0 Array was used. This 
micro-array contains 54675 oligonucteotids (probesets) representing the expressed human genes for each samples. On the 
microarray one gene is represented by one or more probesets.  Each probeset is built up from 25 mer length 
oligonucleotides, so called probes \cite{probeset}. In the present study probesets are the units of observation. For the 
identification of probesets the Affymetrix IDs were used. The Pearson's product-moment correlation was calculated for each 
probeset-pair expression level, and those which have value greater than 0.88 are used to construct the gene co-expression 
network. This network consists of 5000 nodes and 1201480 undirected edges. Nodes represent probeset denoting genes, and 
edges denote their co-expression levels.

\begin{figure} 
\includegraphics[width=5cm]{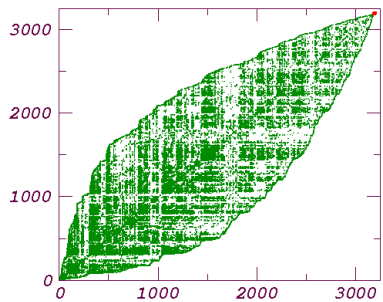} 
\caption{Adjacency matrix of the largest connected 
component of the Gene co-expression network with the threshold value of $\sim 0.89$. Nodes forming largest connecting 
cluster are renumbered in the sequential order for a clear visualization. } 
\label{Nodes} 
\end{figure}

\begin{figure} 
\includegraphics[width=5cm]{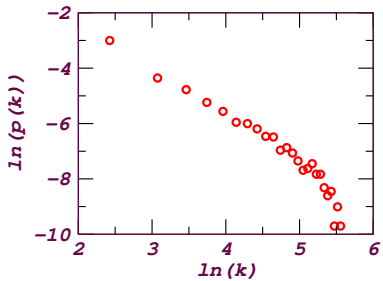} 
\caption{Degree distribution of the largest connected part of the Gene 
co-expression network for threshold 0.89.} \label{degree} 
\end{figure}

From this weighted network, we construct a sparse binary network as following. We choose the value of threshold being 
$r=0.89$, if the co-expression strength is greater than $r$ than the corresponding element in the matrix gets value $1$, 
otherwise $0$. Threshold value of $r=0.89$ leads to a network with much less number of edges, and results into many 
disconnected component. Note that choosing the threshold value is a crucial step and different schemes have been proposed 
to select it \cite{threshold,threshold-rmt}. We sort out the nodes and edges forming largest connecting cluster, which is 
of the size $N=3179$ and $46033$ connections. The average degree of this network is $<k>\sim 30$. RMT analysis is done for 
this biggest component. Fig.~\ref{Nodes} shows the adjacency matrix of this component and Fig.~\ref{degree} is the degree 
distribution. 

\section{Results}
In the following, we present the various RMT results for gene co-expression network constructed above. We calculate the 
eigenvalues and eigenvectors of the adjacency matrix corresponding to the largest connected network. Since this is an 
undirected network, eigenvalues of adjacency matrix are real, and we denote them as $\lambda_i, i=1 \hdots N$. 
Eigenvectors are denoted as $u^k, k=1 \hdots N$.

\subsection{Spacing distribution and $\Delta_3$ analysis}
\begin{figure}
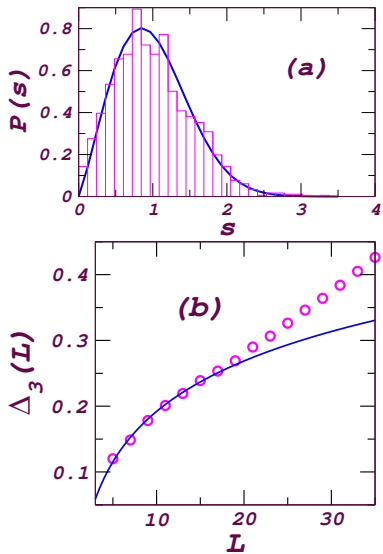

\centerline{\includegraphics[width=5cm]{Spac.eps}}
\centerline{\includegraphics[width=5cm]{Delta3_rev.eps}}
\caption{(Color online) Spacing distribution (a) and $\Delta_3 (L)$
statistics (b) for the eigenvalue spectra of the gene co-expression
network. The histogram in (a) corresponds to the numerical values
and solid line is GOE prediction (\ref{eq-goe})
 of RMT. The circles in (b) are numerical results
(\ref{eq-delta3}) and the solid curve is GOE prediction
(\ref{eq-delta3-goe}) of $\Delta_3$.} \label{fig-Spac}
\end{figure}
From this spectrum we calculate NNSD $P(s)$ as described in the section \ref{RMT} and $\Delta_3(L)$ statistic using 
Eq.~(\ref{eq-delta3}). Fig.~\ref{fig-Spac}(a) shows that NNSD agrees well with the NNSD of GOE matrices (\ref{eq-goe}) 
with the value of Brody parameter \cite{pre2007a,Brody} $\beta \sim 1$.

Fig.~\ref{fig-Spac}(b) plots the $\Delta_3(L)$ statistics. It can be seen that $\Delta_3(L)$ statistic agrees well with 
the GOE statistics up to the value of $L \sim 25$, (which is much less than the same for the corresponding random and 
scale free model networks \cite{pre2007a}). According to the RMT, this implies that besides randomness, the network has 
some specific features. Note that the points which deviate from GOE statistics ($L > 20$), as shown in the 
Fig.~\ref{fig-Spac}(b) can also be analyzed using deformed GOE statistics as shown in \cite{pre2009a}.

\subsection{Eigenvector analysis}
Having calculated spacing distribution and $\Delta_3$ statistics, now we use eigenvector analysis to study the factors 
responsible for the deviation from RMT. We calculate IPR and entropy for all the eigenvectors. The eigenvectors, whose IPR 
and entropy deviate from the random matrix predictions, carry the relevant information. The nodes corresponding to the top 
contributing components of these vectors may be {\it important nodes} in terms of functionality of the whole network. In 
the following we present the Eigenvectors analysis results for the gene co-expression network.

\begin{figure}
\includegraphics[height=7cm,width=0.9 \columnwidth]{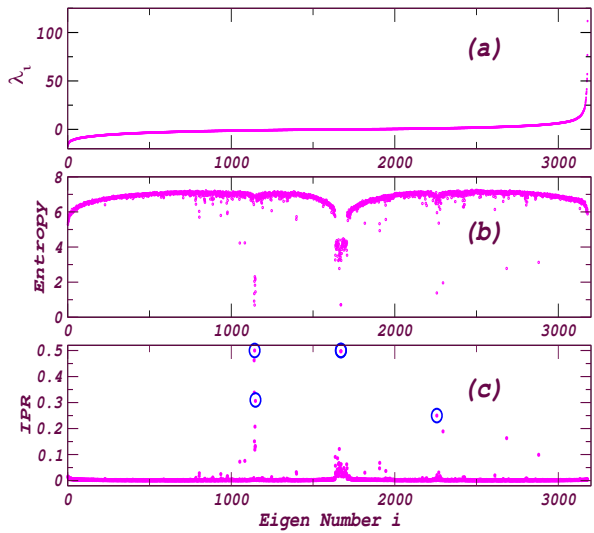}
\caption{(Color online) Eigenvalues (a), entropy (b), and IPR (c) as
a function of eigen number for the threshold value of 0.89. Open
blue circles in (c) correspond to the localized eigenvectors whose
top contributing nodes are listed in the Table~\ref{table1}}
\label{fig-Ent-IPR}
\end{figure}

Fig.~\ref{fig-Ent-IPR}(a) shows eigenvalues in the increasing order. Apart from distinguishably seen high eigenvalues 
towards the end of the spectra, there is a flat part around the zero eigenvalue. Real world networks, in general, are very 
sparse and are reported to have large number of $zero$ eigenvalues \cite{Mendes,Aguiar}. Though for the network we 
consider here, out of 3179 eigenvalues, only approximately 73 ($\sim 2.5\%$ of all eigenvalues) are degenerate with the 
value $zero$. The degeneracy at zero eigenvalue is lesser than many other real world networks \cite{pre2007a}. There are 
nearly 3106 non-degenerate eigenvalues, which could be taken as the effective dimensionality of the network.

We also calculate Shannon entropy for all the eigenvectors using Eq.~(\ref{eq-entropy}), and compare them with those of 
the random vectors. Fig.~\ref{fig-Ent-IPR}(b) shows the entropy as a function of eigen numbers. According to RMT, Shannon 
entropy of a random vector of dimension $N=3106$ is $\ln(3106/2) \simeq 7.35$. Furthermore, RMT predicted value for 
Shannon entropy of a random vector of dimension $N=73$ (corresponding to degenerate part) is $\ln(73/2) \simeq 3.6$. Based 
on these calculations, we can divide eigenvalues into three sets; (A) The non-degenerate part that follows RMT. (B) The 
non-degenerate part, at both ends and at intermediate eigenvalues, which deviate from RMT and expected to contain 
information about {\it important nodes} in the network. (C) The degenerate part with $zero$ eigenvalue, 1636 to 1708 which 
fluctuates around RMT predicted value.

Furthermore, we calculate IPR of all the eigenvectors using Eq.~(\ref{eq-IPR}), and plot in Fig.~\ref{fig-Ent-IPR} (c). It 
shows that IPR of several eigenvalues are localized. For example, vectors corresponding to the 1140 to 1148 eigenvalues 
have $I^k \ge 0.1$, showing that few components contribute more than the other components. Following we enlist some 
localized eigenvectors corresponding to non-degenerate eigenvalues from set (B): $u^{1143}$ (with $I^k \sim 0.5$), 
$u^{1148}$ (with $I^k \sim 0.31$), $u^{2257}$ (with $I^k = 0.25$). Some of the localized eigenvectors corresponding to 
zero eigenvalues are (set (C)); $u^{1636}$ (with $I^k = 0.1$), $u^{1670}$ and $u^{1671}$ (with $I^k \sim 0.5$). We next 
analyze the significant contributors of eigenvectors deviating from the RMT predictions.  The eigenvector $u^{1143}$ 
contains approximately $1/IPR^{1143} = 20$ significant participants. Table~\ref{table1} presents top 5 significant 
contributors (nodes) corresponding to the localized eigenvector mentioned above. Note that original gene number are 
written as in the datasets \cite{data_gene}.
\begin{table}
\resizebox{\columnwidth}{!}{
\begin{tabular}{lccccccr}
\hline
\vline & & Set B & &\vline & Set C & & \vline\\
\hline
\vline & $u^{1143}$ & $u^{1148}$ & $u^{2257}$ & \vline & $u^{1670}$ & $u^{1671}$ & \vline\\
\vline & 202060$\_$at & 227636$\_$at & 202916s$\_$at & \vline & 225921$\_$at & 21435x$\_$at & \vline \\
\vline &   217731s$\_$at  &   205003$\_$at   &   226832$\_$at   & \vline &   212635$\_$at   &   203034s$\_$at   & \vline \\
\vline &   201121s$\_$at   &   211940x$\_$at   &   209860s$\_$at   & \vline &   208645s$\_$at   &   200673$\_$at    & \vline \\
\vline &   221775x$\_$at   &   224616$\_$at   &   218175$\_$at   & \vline &   221511x$\_$at   &   221471$\_$at   & \vline \\
\vline &   229630s$\_$at   &   222203s$\_$at   &   221810$\_$at   & \vline &   231896s$\_$at   &   225950$\_$at   & \vline\\
\hline
\end{tabular}}
\caption{Top five largest contributing nodes in localized eigenvectors for network
constructed with the threshold value of 0.89. The nodes are written in the original
gene number as given in the datasets \cite{data_gene}}
\label{table1}
\end{table}
As shown in the Fig.~\ref{degree}, degree distribution of the connected network analyzed above follows a power law with a 
fat tail, which means that few nodes are hubs, and carry the whole network. But random matrix analysis of eigenvetcors 
reveals that all the most contributing nodes listed above have rather small degree. They are all almost towards bottom of 
the power law distribution.

The degree of all the top contributing nodes in the localized eigenvectors are either well below the average degree or 
around the average degree of the network. Gene, assigned with probeset 202060$\_$at, (corresponding to the node $2299$ in 
the renumbered network) which is the first top contributing node corresponding to eigenvector $u^{1143}$, has a degree 
15,the second top contributing node has a degree 17, the third node has a degree 20. Fourth and fifth top contributing 
nodes have degree 9 each. The top five nodes corresponding to $u^{1148}$ have degree 21, 14, 7, 17 and 24. Those are 
corresponding to eigenvector $u^{2257}$ have degree 1, 1, 6, 3 and 1 respectively. The localized eigenvectors 
corresponding to set (c) are $u^{1670}, u^{1671}$, and top five contributing nodes have degree, in sequential order from 
first to the fifth contributing node (see Table~\ref{table1}), ${2, 4, 8, 1, 3}$ and ${10, 9, 23, 14, 2}$ respectively.

Now we change the threshold value to $0.91$, this threshold value leads to $25,000$ connections in the whole network. This 
network has largest connected cluster of size 2,439 and number of connections 22546. 
The average degree of this network is $<k>\sim 20$. Again we renumber the nodes such that 
nodes in the connected component take value from 1 to 2,439, and  
calculate the eigenvalues and eigenvectors of the adjacency matrix corresponding to this largest connected
network. From the spectrum $NNSD$ and $\Delta_3$ statistics are calculated, and 
these two show similar GOE statistics as shown in Fig.\ref{fig-Spac} for $r=0.89$.

\begin{figure}
\includegraphics[height=7cm,width=0.9 \columnwidth]{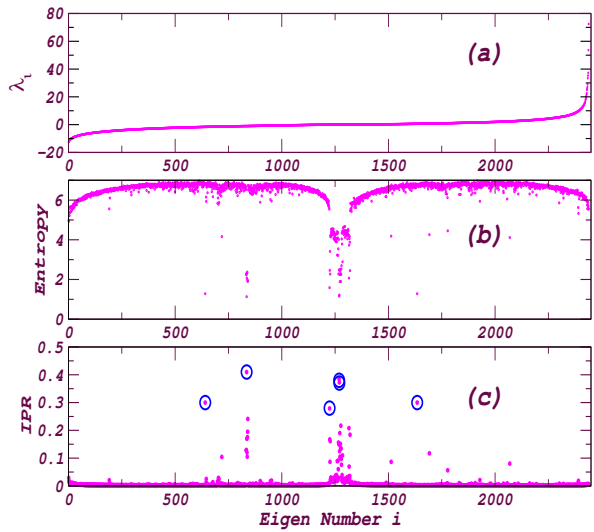}
\caption{(Color online) Same as Fig.4 but for threshold value of
0.91.  Open blue circles correspond to localized eigenvectors whose
top five contributing nodes are presented in the
Table~\ref{table2}.} \label{fig-Thr2-Ent-IPR}
\end{figure}

Fig.~\ref{fig-Thr2-Ent-IPR} plots eigenvalues (a), entropy 
(b) and IPR (c) as a function of eigen number. Entropy and IPR are calculated using Eq.~(\ref{eq-entropy}) and 
(\ref{eq-IPR}) respectively. Out of 2,439 eigenvalues, approximately 96 are degenerate with the value $zero$. It means 
that there are nearly 2343 non-degenerate eigenvalues, which could be taken as the effective dimensionality of the 
network. According to RMT, Shannon entropy of a random vector of dimension $N=2343$ is $\ln(2343/2) \simeq 7.0$. On the 
other hand, RMT predicted value for Shannon entropy for degenerate eigenvectors is $\ln(96/2) \simeq 3.9$. Based on these 
calculations, again we can divide eigenvalues in three sets (A), (B) and (C). Localized eigenvectors corresponding to 
non-degenerate part are: $u^{835}$(IPR=0.41), $u^{1635}$ (IPR=0.3), $u^{641}$(IPR=0.3), $u^{840}$ and $u^{841}$ (with 
$\lambda=1$, IPR=0.195 and 0.24) Localized eigenstates corresponding to zero eigenvalues (set (c)) are: $u^{1269}$ 
(IPR=0.38), $u^{1270}$ (IPR=0.37), $u^{1224}$ (IPR=0.28). Significant contributors in localized eigenvectors are written 
in Table~\ref{table2}.

\begin{table}
\resizebox{\columnwidth}{!}{
\begin{tabular}{lcccccccr }
\hline
\vline & & Set B & & \vline & & Set C & & \vline \\
\hline
\vline & $u^{835}$ & $u^{1635}$ & $u^{641}$ & \vline & $u^{1269}$ & $u^{1270}$ & $u^{1224}$ & \vline \\
\vline &  210338s$\_$at   &  208666s$\_$at   &  201121s$\_$at   & \vline &  211733x$\_$at  &  201494$\_$at  &  230416$\_$at   &\vline \\
\vline &  210418s$\_$at   &  224819$\_$at   &  208667s$\_$at   & \vline &  230869$\_$at   &  223209s$\_$at  &  228283$\_$at  &\vline \\
\vline &  202178$\_$at   &  209460$\_$at   &  223716$\_$s$\_$at   & \vline &  228045$\_$at   &  225284$\_$at   &  238494$\_$at  &\vline \\
\vline &  38398$\_$at  &  226395$\_$at   &  224644$\_$at   & \vline &  211733x$\_$at  &  201494$\_$at   &  230416$\_$at  &\vline \\
\vline &  213347x$\_$at   &  201525$\_$at   &  200626s$\_$at  & \vline &  242317$\_$at  &  212788x$\_$at   &  212474$\_$at   &\vline\\
\hline
\end{tabular}
}
\caption{Top contributing nodes (genes) in the localized eigenvectors for the threshold value 0.91}
\label{table2}
\end{table}
The degree distribution of the largest component at this threshold follows a power law as well, revealing the scalefree 
nature of this component. Increasing threshold preserves scalefree property of the network. Some nodes are hubs which carry 
the whole network and enjoy the structural importance. Again we find that the top contributing nodes are not the ones with 
very high degree. For two different threshold values Tables \ref{table1} and \ref{table2} show the largest contributing 
co-expressing genes in the corresponding localized eigenvectors. We find that choosing threshold is very important for the 
analysis of Gene co-expression networks, as we can see that top five largest contributing nodes differ entirely (except 
one) as threshold value is changed. This suggests that, though the gross structure of whole network (Fig.~\ref{Nodes}) and 
scale-free 
property, remains unchanged, value of threshold has a strong effect on the network leading to entirely different sets 
(except few) of largest contributing nodes for two different threshold values. Appendix enlists the genenames 
corresponding to the probesets identifiers as given in \ref{table1} and \ref{table2}.

\section{Conclusions and Discussions}
Using RMT, we have analyzed gene co-expression network constructed by applying two different threshold values to the data 
obtained from six brain regions that are metabolically relevant to Alzheimer's disease \cite{data_gene}. The NNSD of 
adjacency matrix of the largest connecting component of the network follows universal GOE statistics (with $\beta \sim 
1$). This universality adds one more feature, based on the spectral correlations, to the gene co-expression network which 
is common with different model networks \cite{pre2007a} proposed to capture various structural properties of real world 
networks.

The NNSD gives information about the short range correlations among the eigenvalues. To probe the long range correlations 
we have studied spectral rigidity via $\Delta_3 (L)$ statistics. This analysis shows that the gene co-expression network 
considered here follows RMT prediction of GOE for very long range of $L$. Beyond this value of $L$ deviation in the 
spectral rigidity is seen, indicating a possible breakdown of universality. This means the network under consideration has 
{\it sufficient} randomness {\it which may due to robustness of the systems}, with regularity {\it which may be to perform 
some functional task}. Mixture of random connections and regular structure have been emphasized at various places, for 
instance information processing in the brain is considered to be random connections among different modular structure 
\cite{face}.

Deviation from the universal RMT predictions identify system-specific, non-random properties of system under 
consideration, might provide clues about important interactions. To extract these system dependent information we have 
performed eigenvector analysis. This analysis reveals that there are some eigenvectors which are highly localized. The 
component $l$ of a given eigenvector relates to the contribution of node (corresponding gene) $l$ to that eigenvector. 
Hence, the distribution of the components contains information about the number of genes contributing to a specific 
eigenvector. Inverse participation ratio IPR, as defined in Eq.~(\ref{eq-IPR}) , distinguishes between one eigenvector 
with approximately equal components and another with a small number of large components. According to the RMT predictions, 
the largest contributing nodes (genes) in the localized eigenvectors may have important function, or important functional 
relations among them.

The largest connected component is scale-free indicating the structural importance of few nodes (hubs). Eigenvector 
analysis shows that top contributing nodes in the localized eigenvectors have relatively low degrees. Note that genes 
which are hubs or those which connect different communities are also important, as shown by several earlier studies in the 
network framework \cite{rev-network,modular}, but the aim of the present work is look for the important genes beyond these 
structural measures. Changing the value of threshold, while keeping the scale-free structure of network same, has drastic 
impact on the localization property of eigenvectors. All most all the top contributing nodes differ for two different 
threshold value, indicating impact on the global properties of the underlying network.

Last, we discuss here the importance of the analysis and future implications of the results presented in the paper. 
Several studies have shown that the development of multi-target drugs might give better results than the traditional 
methods targeting a single protein. Single target-design might not always give satisfactory results, as there might be a 
backup system, which replaces the function of the inhibited target protein. By using multi-target drugs one can decrease 
the functionality of entire protein cascades producing more effective results. For example, studies have shown that aging 
is strongly linked with age-related diseases, and they share a common signaling network. Signaling hubs of the age-related 
protein-protein interaction subnetwork may be good candidates for age-related drug-targets. Multi-target drugs attacking 
hubs of the protein-protein interaction network, 'hub-links' (links connecting hubs), bridges (inter-modular links having 
high 'betweenness centrality') or nodes in the overlap of numerous network modules, might give better results 
\cite{multi-target,review-multi-target}. Similarly, targeting genes corresponding to the largest contributing nodes in 
localized eigenvectors may lead to important effect as well. Future investigations are sought in order to know the 
functionality of these genes corresponding to the top contributing nodes in the localized eigenvectors, which could be 
then used for such multi-target drug designs.

\appendix*
\section{}
Tables \ref{table3} and \ref{table4} correspond to probesets identifiers from tables \ref{table1} and \ref{table2} 
respectively. First column of these tables are probeset identifiers (Affymetric ID) and second column dictates the 
corresponding genenames. However, the he function of some transcripts is not known yet, and some of them has no gene name. 
The value '-' in the gene name column indicates that information is not available. Note that there are many reasons for 
probesets without detailed annotation. We know the sequence on microarray for each probesets. On the chip we get all 
expressed genes, but we do not have secure info for all the gene functions. As the knowledge is growing with the latest 
available technologies, this gap is decreasing with time. One sure information for the probeset is the Affymetric ID as 
given in the table I and II \cite{probeset}.
\begin{table}
\resizebox{\columnwidth}{!}{
\begin{tabular}{lcccccccr}
\hline
\vline  & Probeset  & Gene name  & \vline \\
\hline
\vline &  202060$\_$at &   Ctr9, Paf1/RNA polymerase II  & \vline\\
\vline &  227636$\_$at &  - & \vline\\
\vline &  202916s$\_$at & family with sequence similarity 20, member B& \vline\\
\vline &  225921$\_$at &  ninein (GSK3B interacting protein)& \vline\\
\vline &  214351x$\_$at & ribosomal protein L13 & \vline \\
\vline &  217731s$\_$at & integral membrane protein 2B &\vline\\
\vline &  205003$\_$at & dedicator of cytokinesis 4 & \vline \\
\vline &  226832$\_$at & - &\vline \\
\vline &  212635$\_$at & transportin 1 & \vline\\
\vline&  203034s$\_$at & ribosomal protein L27a & \vline\\
\vline &  201121s$\_$at & progesterone receptor membrane component 1 &\vline\\
\vline &  211940x$\_$at & - &\vline \\
\vline &  209860s$\_$at & annexin A7 &\vline \\
\vline &  208645s$\_$at & ribosomal protein S14 &\vline\\
\vline &  200673$\_$at & lysosomal protein transmembrane 4 alpha &\vline\\
\vline &  221775x$\_$at & ribosomal protein L22 & \vline\\
\vline &  224616$\_$at & dynein, cytoplasmic 1 & \vline \\
\vline &  218175$\_$at & coiled-coil domain containing 92 & \vline \\
\vline &  221511x$\_$at & cell cycle progression 1 & \vline \\
\vline &  221471$\_$at & serine incorporator 3 & \vline \\
\vline &  229630s$\_$at & Wilms tumor 1 associated protein & \vline \\
\vline &  222203s$\_$at & retinol dehydrogenase 14 & \vline \\
\vline &  221810$\_$at & RAB15, member RAS onocogene family & \vline \\
\vline &  231896s$\_$at & density-regulated protein & \vline\\
\vline &  225950$\_$at & - & \vline\\
\hline
\end{tabular}
}
\caption{Genenames corresponding to the probesets for the threshold value 0.89}
\label{table3}
\end{table}
\begin{table}
\resizebox{\columnwidth}{!}{
\begin{tabular}{lcccccccr}
\hline
\vline  & Probeset  & Gene name  & \vline \\
\hline
\vline & 210338s$\_$at & heat shock 70kDa protein 8 &\vline\\
\vline & 208666s$\_$at & suppression of tumorigenicity 13 &\vline\\
\vline & 201121s$\_$at & progesterone receptor membrane component 1 &\vline \\
\vline &211733x$\_$at & sterol carrier protein 2 &\vline\\
\vline & 201494$\_$at & prolylcarboxypeptidase & \vline\\
\vline &230416$\_$at & - &\vline\\
\vline & 210418s$\_$at & isocitrate dehydrogenase 3 (NAD+)  &\vline\\
\vline & 224819$\_$at & transcription elongation factor A (SII) &\vline\\
\vline & 208667s$\_$at & suppression of tumorigenicity 13& \vline\\
\vline & 230869$\_$at & family with sequence similarity 155 & \vline\\
\vline & 223209s$\_$at & selenoprotein S &\vline\\
\vline & 228283$\_$at & COX assembly mitochondrial protein homolog &\vline\\
\vline & 202178$\_$at & protein kinase C, zeta &\vline\\
\vline & 209460$\_$at & 4-aminobutyrate aminotransferase &\vline\\
\vline & 223716s$\_$at & zinc finger, RAN-binding domain & \vline\\
\vline& 228045$\_$at & - &\vline\\
\vline& 225284$\_$at & DnaJ (Hsp40) homolog, subfamily C &\vline\\
\vline& 238494$\_$at &TNF receptor-associated factor 3 & \vline\\
\vline& 38398$\_$at & MAP-kinase activating death domain &\vline\\
\vline& 226395$\_$at & hook homolog 3 (Drosophila)&\vline\\
\vline& 224644$\_$at & - &\vline\\
\vline& 211733x$\_$at & sterol carrier protein 2 &\vline \\
\vline& 201494$\_$at & prolylcarboxypeptidase & \vline\\
\vline& 230416$\_$at & - & \vline \\
\vline& 213347x$\_$at & ribosomal protein S4, X-linked & \vline \\
\vline& 201535$\_$at & ubiquitin-like 3 &\vline\\
\vline& 200626s$\_$at & martin 3&\vline\\
\vline&242317$\_$at & HIG1 hypoxia inducible domain family &\vline\\
\vline&212788x$\_$at & ferritin, light polypeptide & \vline\\
\vline& 212474$\_$at & AVL9 homolog (S. cerevisiase) & \vline\\
\hline
\end{tabular}}
\caption{Genenames corresponding to the probesets for the threshold value 0.91}
\label{table4}
\end{table}

\end{document}